# Moiré-Engineered Excitonic Landscape and Phonon-Mediated Recombination in Twisted $WSe_2$ Bilayers

*Memansa Thapa[1], Aksa Thomas[1], Jayalekshmi U. J.[2,3], Krishna Prasad Bera[4], Darshit Solanki[4], Kenji Watanabe[5], Takashi Taniguchi[5], Ajay Kumar Shukla[2,3], Anindya Das[4], and Ajay Soni[1]**

[1]School of Physical Sciences, Indian Institute of Technology Mandi, Mandi 175005, Himachal Pradesh, India
[2]CSIR-National Physical Laboratory, Dr. K. S. Krishnan Marg, New Delhi, 110012, India
[3]Academy of Scientific and Innovative Research (AcSIR), Ghaziabad, 201002, India
[4]Department of Physics, Indian Institute of Science, Bangalore 560012, Karnataka, India
[5]National Institute of Material Science, 1-1 Namiki, Tsukuba 305-0044, Japan
[*]*Email for Correspondence: ajay@iitmandi.ac.in*

**Abstract:** Twist-angle engineering in two-dimensional (2D) van der Waals (vdW) heterostructures unlocks an additional degree of freedom to tune optoelectronic properties and probe multibody quantum interactions. Here, we report light emission from the moiré superlattice of a twisted bilayer of tungsten diselenide ($WSe_2/WSe_2$) encapsulated in insulating hexagonal boron nitride (hBN). The low-temperature photoluminescence (PL) spectroscopy reveals signatures of moiré-potential induced strong interlayer excitonic emission and phonon-assisted recombinations, while the twisting significantly suppresses the emission from localized defect-bound excitons. The moiré potential redistributes carriers into indirect valleys, thereby enhancing recombination efficiency and stabilizing the interlayer excitons. Our findings establish that precise control of twist angle and dielectric environment provides a new route for engineering excitonic systems for exploring exciton-phonon interactions and associated quantum phenomena in transition metal dichalcogenides.

## Introduction

Engineering the physical properties on demand is an important strategy for rational designing and development of technologically advanced materials from the library of existing materials.[1-3] Semiconducting 2D-vdW are class of materials, which have been extensively studied for next-generation quantum optoelectronics using light-matter interactions. These materials exhibit a thickness tunable electronic band gap along with strong photo-responses.[4-12] The discovery of superconductivity in the twisted graphene layers have opened a new era of twistronics and following that a wide range 2D transition metal dichalcogenides (TMDs) have been explored for both technological advancements and fundamental insights.[1,13,14] Being direct band gap material in the monolayer limit, TMDs serve as an important test bed for understanding fundamental optical processes of exciton dynamics and quantum interactions at atomic level. Furthermore, twisting of vdW atomically thin layers has emerged as a powerful strategy to engineer artificial heterostructures with tunable physical properties, enabling advances in optoelectronics and photonic quantum technologies. The mechanical exfoliations of the graphene, and TMDs has been a versatile methodology for the on demand fabrication of heterostructures.[15,16] The TMDs have covalently bound atomically thin layers held together by vdW forces, and this geometric flexibility allows the stacking of similar or distinct layers and offering an innovative platform to explore a broad range of nascent quantum phenomena, such as superconductivity, orbital magnetism, mott-insulating states, and exciting optical properties.[7,17-20] Importantly, there are notable changes to the band structure and optical properties of TMDs for instance, the quenching of bright PL upon changing the layer configuration from monolayer to bilayer owing to the direct to indirect bandgap transition.[21] Additionally, the twisting results in moiré superlattices, which reshapes the atomic arrangement and modifies the electronic structure as well as optical selection rules.[22,23] Quan *et al*. have reported the renormalization of phonon spectra in a twisted $MoS_2$ homobilayer due to local strain originating from different stacking configurations.[18] Additionally, the impact of the twist angle on phonon hybridization of optical phonons in twisted $WSe_2$ bilayers due to the moiré-potential has been studied by Bera *et al.*[24] However, the impact of the twist angle on exciton localization in TMDs due to the moiré-potential is scarcely reported in the literature.

In a monolayer or bilayer $WSe_2$, the PL spectrum is composed of direct and indirect excitons, charged excitons[8,9,16,25] biexcitons and excitons bound to local vacancies as well as defects.[10,17,26,27] In this study, we demonstrate the emergence of moiré superlattice-induced excitonic complexes in twisted bilayers formed by stacking two $WSe_2$ monolayers at a finite twist angle and encapsulating them in hBN. From the temperature-dependent Raman and PL measurements on monolayer (ML), natural bilayer (NBL), artificial bilayer with twist angle ~ 60º (ABL), and twisted bilayer with twist angle ~ 2º (TBL) $WSe_2$, we show that precise control over the twist angle enables the realization of impeccable tunable phononic and excitonic properties. These systems exhibit pronounced phonon-assisted interlayer states, providing a versatile platform to explore phonon-mediated valley physics and interlayer exciton dynamics, with potential implications for twist-angle-controlled, phonon-engineered valleytronic and excitonic devices.

**Results and Discussion**

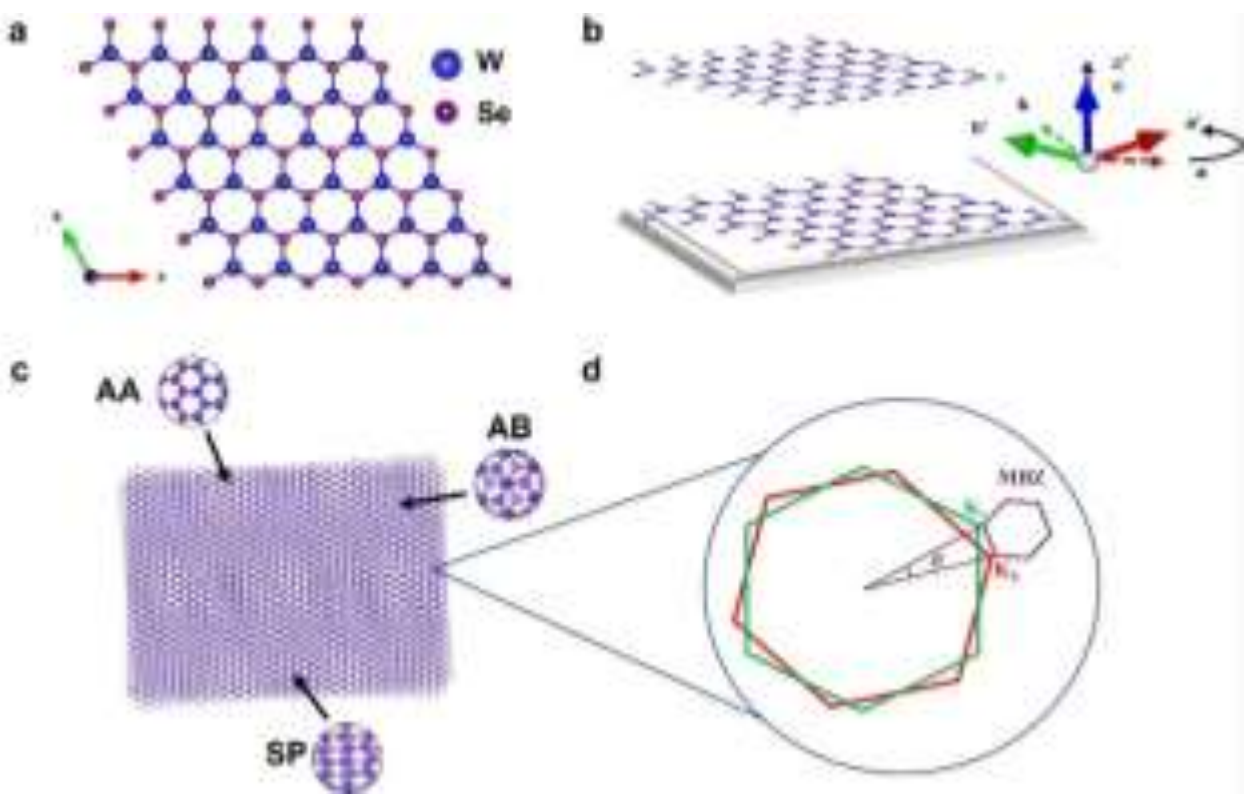

**Figure 1.** Structural configuration of twisted bilayers. (a) Top-view of the crystal structure of 2H-$WSe_2$. (b) Schematic of twisting of bilayer $WSe_2$. (c) Schematic representation of the moiré superlattice resulting from the twisting of two $WSe_2$ monolayers, with insets showing the various atomic stackings AA, AB, and SP (saddle point). (d) The Brillouin zones of top (green) and bottom (red) monolayers, and the mini-Brillouin zone (MBZ) formed due to the interlayer twisting (purple) of the monolayers.

Figure 1 illustrates the structural configuration, schematics for the fabrication of twisted bilayers, and formation of moiré superlattice along with modified Brillouin zone. Figure 1a shows the top-view of 2H-$WSe_2$ having a hexagonal arrangement of tungsten (W) and selenium (Se) atoms. Figure 1b shows the schematic of the fabrication of rotationally aligned twisted bilayer by a dry-transfer technique of mechanically exfoliated monolayers. Further details of sample preparation are provided in the Experimental Section. The precise stacking of the twisted layers gives a long-range moiré superlattice (Figure 1c), having a distinct periodicity, given by $a_M = \frac{a}{2\,sin\left(\frac{\theta}{2}\right)}$, where '$a$' is the lattice constant of 2H-$WSe_2$, and $\theta$ is the relative twist angle between monolayers. Depending upon the local stacking arrangement, the moiré superlattice creates high symmetry regions, such as AA, AB/BA stacking, along with an intermediate configuration between AA and AB/BA, described as SP (saddle point).[17,18,24,28,29] Here, the AA stacking has the highest potential energy, whereas AB/BA represents the minimum energy configuration. On the other hand, for the twisted bilayers, the extent of atomic reconstruction depends on the strength of the underlying moiré potential arising from the interlayer coupling in twisted configuration.[30,31] The corresponding Brillouin zones (BZ) of the upper (green) and lower (red) monolayers (Figure 1d), along with the emergent BZ, demonstrate how the interlayer twisting leads to the formation of a reduced mini-Brillouin zone (MBZ). The breaking of inversion symmetry in the MBZ is associated with a unique electronic structure, which is expected to modify the optical responses of the twisted bilayer.

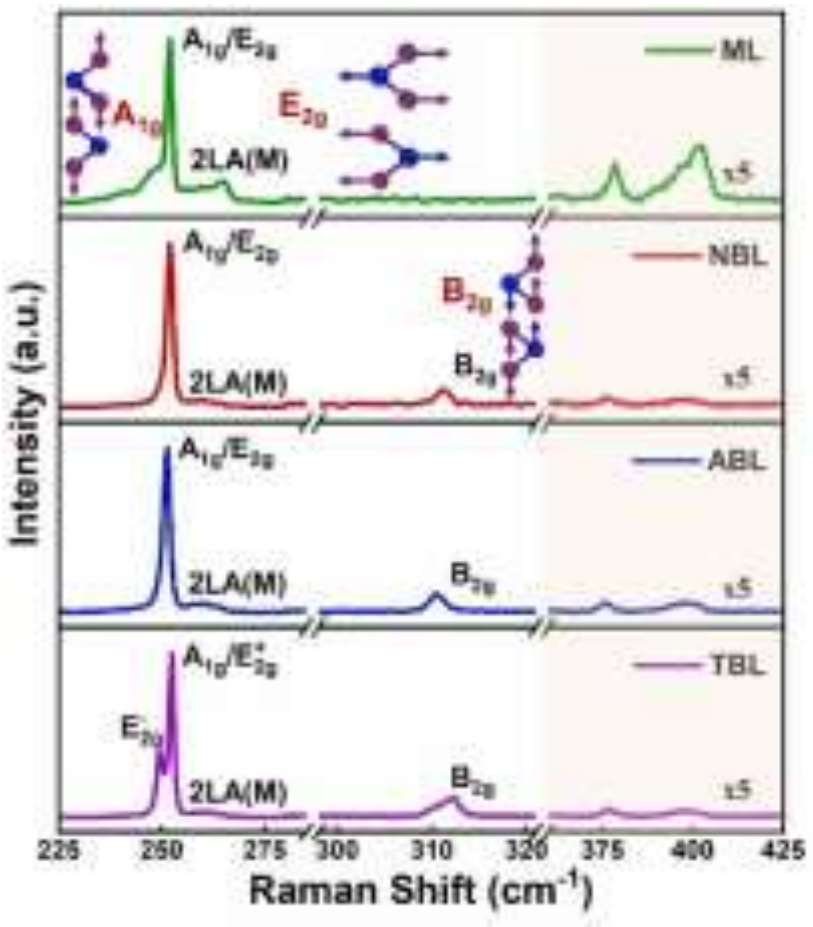

**Figure 2.** Raman spectra at 3 K for various stackings of $WSe_2$ layers. Raman spectra of monolayer (ML), natural bilayer (NBL), and fabricated bilayers with twist angle, $\theta \sim 60^o$ (artificial bilayer, ABL) and $\theta \sim 2^o$ (twisted bilayer, TBL). The insets illustrate the schematics of atomic displacements of the $A_{1g}$, $E_{2g}$ and $B_{2g}$ modes, where blue and purple spheres denote W and Se atoms, respectively. The intensity of high frequency modes (370 - 425 $cm^{-1}$) is amplified ~ 5 times for better visibility.

The Raman spectra, at 3 K, for various stacking configurations, ML, NBL, ABL and TBL is shown in Figure 2, where the characteristic Raman modes and corresponding atomic displacement schematics for $A_{1g}$, $E_{2g}$ and $B_{2g}$ symmetries are presented in the insets.[24] Here, the $A_{1g}$ modes have the out-of-plane vibrations of the Se atoms while the W atoms are at the center of the mass, on the other hand, the $E_{2g}$ modes have in-plane vibrations of both the atoms. Generally, the intense peak around ~ 250 cm$^{-1}$ has the dependence on the excitation energy and is well reported in the literature for having the contributions from both $A_{1g}$ and $E_{2g}$ modes.[24,32-34] In the present case of excitation with ~ 532 nm laser, the $A_{1g}$ mode is dominant. In contrast to NBL and ABL, the Raman spectrum of TBL shows splitting of the ~ 250 cm$^{-1}$ peak. These results were consistent with prior theoretical and experimental findings by Bera *et al.*[24], where the splitting is attributed to the phonon hybridization and the moiré-potential induced lattice reconstruction due to strong interlayer coupling. The weak mode around ~ 260 cm$^{-1}$ is the second-order longitudinal acoustic phonon mode, (2LA (M)).[24] The mode around ~ 310 cm$^{-1}$ is a nondegenerate interlayer coupling mode, $B_{2g}$, which is present in all bilayers and evidently absent for ML $WSe_2$. The higher-order modes between 350 - 450 cm$^{-1}$ appears with very poor intensity, and thus shown with an amplified intensity (~ 5 times) for clarity. The evolution of phonon modes as a function of temperature supports the renormalization of phonons induced by moiré potential and interlayer hybridization effects.[17,18,24] Details of temperature-dependent Raman spectroscopy measurements are provided in Figure S1. This consistency of the Raman signatures with previously reported literature,[18,24] validates the twist angle configuration in TBL and provides a reliable foundation for further optical measurements.

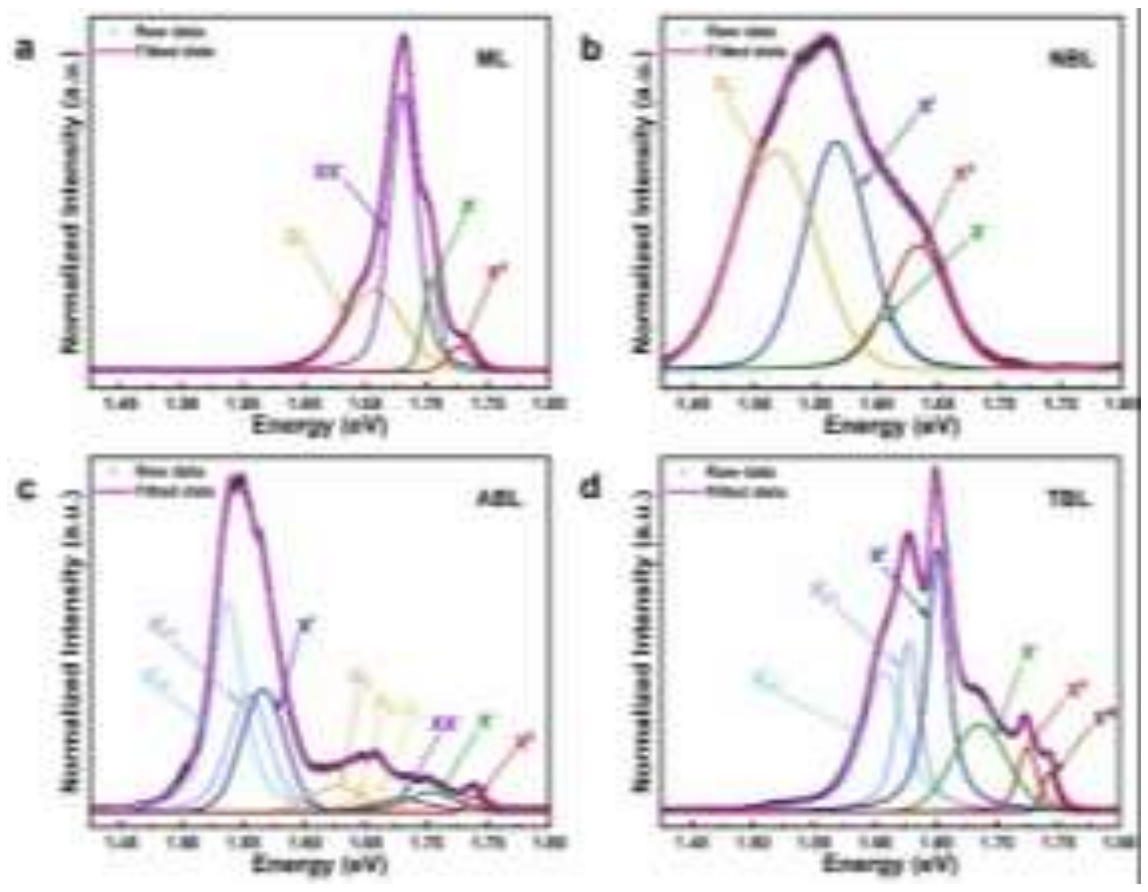

**Figure 3.** Normalized PL spectra at 3 K for various stacking layers of $WSe_2$. (a) Monolayer (ML), (b) natural bilayer (NBL), (c) artificial bilayer fabricated with twist angle, $\theta \sim 60^{\circ}$ (ABL), and (d) twisted bilayer with $\theta \sim 2^{\circ}$ (TBL).

Being a direct bandgap semiconductor, the ML $WSe_2$ exhibits a strong PL intensity, which decreases with the number of layers in multilayers due to the thickness-dependent direct (K-K) to indirect (Q-K) bandgap transition.[35] For bilayers, the conduction band minimum resides at the Q point and the valence band maximum remains at K, which reduces the radiative recombination efficiency, [27,28,36] as shown by the room temperature PL mapping and spectra (Figure S2). To understand the effect of twist angle on the excitonic landscape, we studied the normalized PL spectra at 3 K for all samples (Figure 3).

The PL spectrum of ML (Figure 3a) has emission from the neutral exciton ($X^0$) at ~ 1.73 eV, and three other transitions with energy at ~ 1.70, ~ 1.68 and ~ 1.65 eV described as negatively charged trion ($X^-$), negatively charged biexciton ($XX^-$), and defect-bound ($D_1$) exciton, respectively. The corresponding energy difference for negatively charged trion is, ($X^0 - X^-$) ~ 30 meV, for the negatively charged biexciton is ( $X^0 - XX^-$) ~ 50 meV[16] and for defect-bound exciton is ($X^0 - D1$) ~ 80 meV, respectively. The observations and assignment of excitonic complexes are consistent with prior reports on monolayer $WSe_2$.[7,9,10,25,35-39] Figure 3b presents the PL spectrum of a NBL, which displays $X^0$ (~ 1.63 eV) and $X^-$ (~ 1.60 eV) alongside a pronounced interlayer exciton, ($X^I$ ~ 1.56 eV) and a defect bound exciton ($D_1$ ~ 1.52 eV). The emergence of the $X^I$ is attributed to interlayer band-edge offsets and momentum-indirect recombination in NBL.[4,10,24,27,28,36,40-43] Here, $D_1$ is energetically related to the Se vacancy bound excitons in the range of 80 - 120 meV below $X^0$ .[4,10,24,27,28,41,43] In a ML $WSe_2$, inversion symmetry is inherently broken, leading to strong valley-contrasting optical selection rules and tightly bound excitons. In contrast, for NBL, inversion symmetry is restored and interlayer coupling is strong, which facilitates interlayer hybridization and results in a redshifted indirect excitonic emissions with broader linewidths. The PL spectra of ABL (Figure 3c) and TBL (Figure 3d) show a richer excitonic landscape in comparison to NBL.[10,43,44] For ABL, several defect-bound emission peaks, denoted as $D_1$, $D_2$, $D_3$, have been observed which is expected as the artificial stacking can create vacancy as well as trap defects during the fabrication of the heterostructures, and similar observations have been reported in the literature for NBL and the artificial heterostructures (bilayer made from two monolayers).[10,28,36,41,44] Furthermore, in addition to these primary excitons, we observed two broad and well-resolved PL emissions, lying at ~ 12 meV $\left(X^I_{p1}\right)$ and ~ 28 meV $\left(X^I_{p2}\right)$ below $X^I$. These emissions are coming from the phonon assisted activation of interlayer excitons and named as phonon replicas, where $X^I_{p1}$ involves one phonon while $X^I_{p2}$ has contribution from two phonons, resulting from nearly resonant exciton-phonon scattering processes.[4,43,45] Interestingly, excitonic landscape for TBL is entirely defect-free in the 1.5 - 1.75 eV energy range, indicating strong moiré-potential induced exciton localization (Table S1).[36] Using multiple-Voigt fitting and comparison with the available literature, we identified multiple distinct excitonic emission peaks with emission energies at ~ 1.74 eV ($X^M$), ~ 1.72 eV ($X^0$), ~ 1.69 eV ($X^-$), ~ 1.65 eV ($X^I$), ~ 1.63 eV ($X^I_{p1}$), and ~ 1.61 eV ($X^I_{p2}$). For TBL, atomic reconstruction leads to the formation of moiré superlattices with spatially varying stackings (AA, AB/BA, and SP), which breaks the inversion symmetry while reducing the effective interlayer coupling compared to the NBL. At higher energy stacking configuration, the excitons become spatially localized within the moiré potential, which results in enhancement of effective band gap.[7,46] Additionally, the blueshift of the interlayer excitonic emissions as well as significantly narrower linewidth relative to the NBL, suggest a prolonged excitonic lifetime.[47,48] In $WSe_2$, the excited states of exciton exist at ~ 130 meV above $X^0$,[49-51] thus, the observed $X^M$ (~ 19 meV above $X^0$) is expected to be a moiré localized exciton. Since the defect-bound states get populated at low excitation power and saturate with an increase in power, the relative integrated PL intensity (RIPL) for $X^M$ continuously increases, (Figure S3), therefore $X^M$ is not associated with a defect mediated state.[16,52,53] Additionally, similar to ABL, two well-resolved and sharp phonon replicas with a much higher intensity than the primary excitons have been observed at ~ 24 meV $\left(X^I_{p1}\right)$ and ~ 44 meV $\left(X^I_{p2}\right)$ below $X^I$ of TBL. [4,43] The replicas give a direct spectroscopic fingerprint of strong phonon assisted recombination pathways for the interlayer excitons and allow quantitative assessment of phonon energies, coupling strength, and localization effects.[54,55] Interestingly, the energy difference of phonon replicas for TBL matches with the energy of the zone-center $E''$ optical phonons (~ 22 meV),[56-58] which suggests that the replicas are related to the interaction of the interlayer excitons and $E''$ phonons at Γ point. To understand the effect of temperature on excitonic complexes in TBL, we conducted a temperature-dependent PL study from 3 K to 300 K.

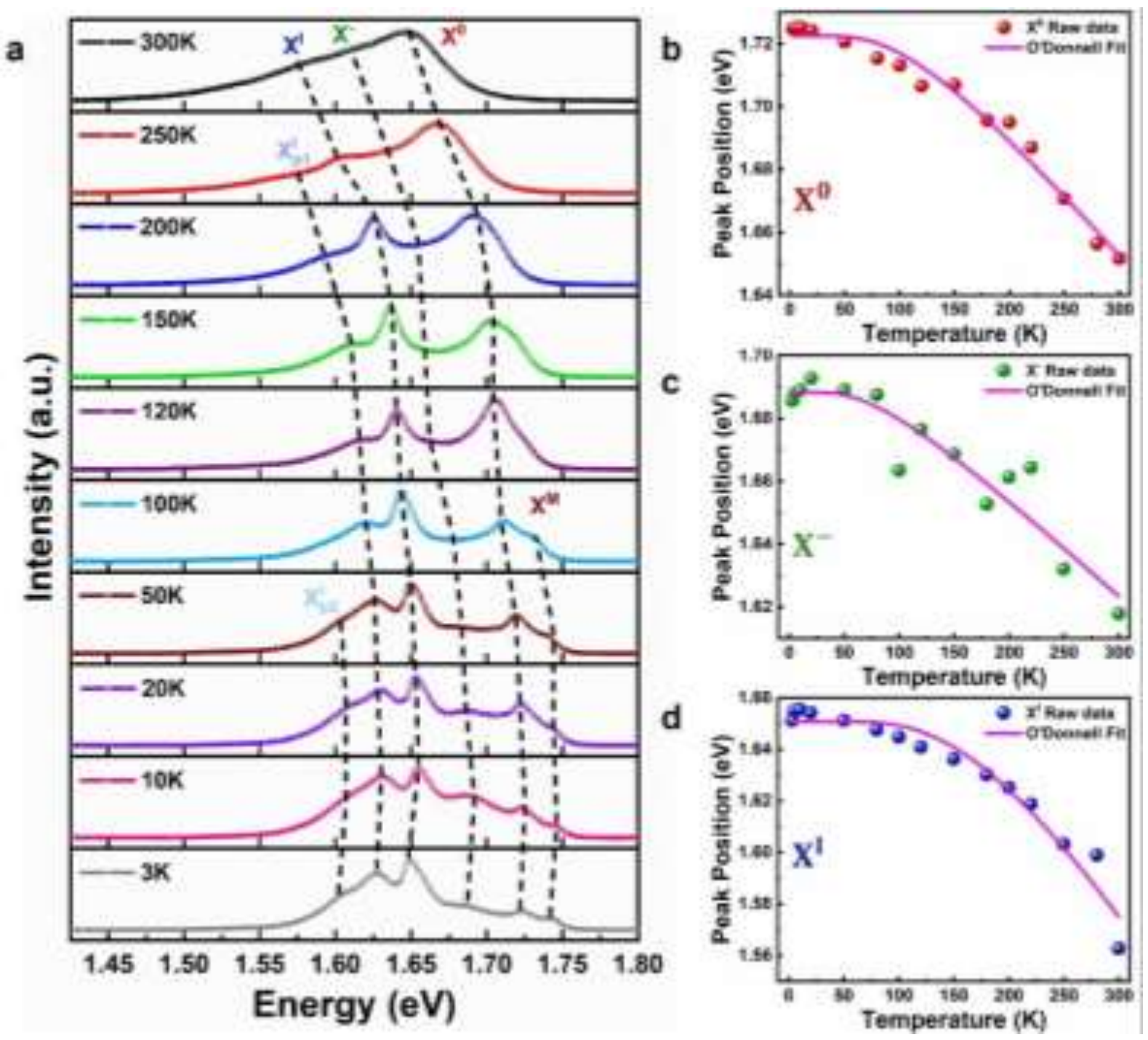

**Figure 4.** Temperature dependence of PL for TBL. (a) Normalized PL spectra of TBL between 3 K to 300 K. Temperature-dependent shifting of (b) neutral exciton ($X^0$), (c) trion ($X^-$), and (d) interlayer exciton ($X^I$), with solid lines representing the O'Donnell equation fit.[8,59]

Figure 4a shows the evolution of the PL spectrum as a function of temperature in the range 3 K to 300 K. Notably, within the temperature range 3 K - 50 K, the phonon replicas are well resolved and prominent, signifying strong exciton-phonon coupling. However, as the temperature rises, the replicas gradually smear their individual prominence and broaden. This merging arises as the population of thermally activated phonons increases with an increase in temperature. Thus, a larger number of phonons causes more scattering, which in turn enhances non-radiative recombination, making the emissions less coherent and broader. Among these replicas, the $X^{I}_{p1}$ fades out above 50 K, probably because it arises from a complex phonon interaction that likely require multiple phonons acting together. The existence of $X^0$, $X^-$ and $X^I$ up to room temperature reflects their large binding energies exceeding the thermal energy $K_BT$ ~ 26 meV at 300 K. Notably, as the temperature rises, the interaction of excitons with the thermally activated phonons allows a redistribution of energy and explains the merging into a broad exciton band at 300 K. To quantify the exciton-phonon interactions across the different radiative channels in TBL, we analyzed the observed redshift of $X^0$, $X^-$, and $X^I$, as shown in Figure 4b-d, using the O'Donnell equation[8,59] $E_g(T) = E_g(0) - S \times E_{ph}\left[coth\left(\frac{E_{ph}}{k_BT}\right) - 1\right]$, where $E_g(0)$ is transition energy at $T = 0$ K, $S$ is exciton-phonon coupling constant, and $E_{ph}$ is average phonon energy involved in exciton-phonon interactions. From the fits, we extract $S$ ~ 2.39, $E_{ph}$ ~ 27.4 meV for $X^0$, $S$ ~ 1.88, $E_{ph}$ ~ 19.7 meV for $X^-$, and $S$ ~ 4.27, $E_{ph}$ ~ 48.2 meV for $X^I$. The larger values of $S$ and $E_{ph}$ for $X^I$ indicates a stronger interaction with phonons, consistent with their momentum-indirect nature and interlayer character, which further confirms the strong exciton-phonon interactions in TBL. Additionally, the change in the bandgap due to thermal expansion of the lattice and its relevance to the Debye temperature have been modelled using the Varshni equation, as shown in Figure S4-S6, further the fitting parameters extracted from the fittings are tabulated in Table S2 and S3.

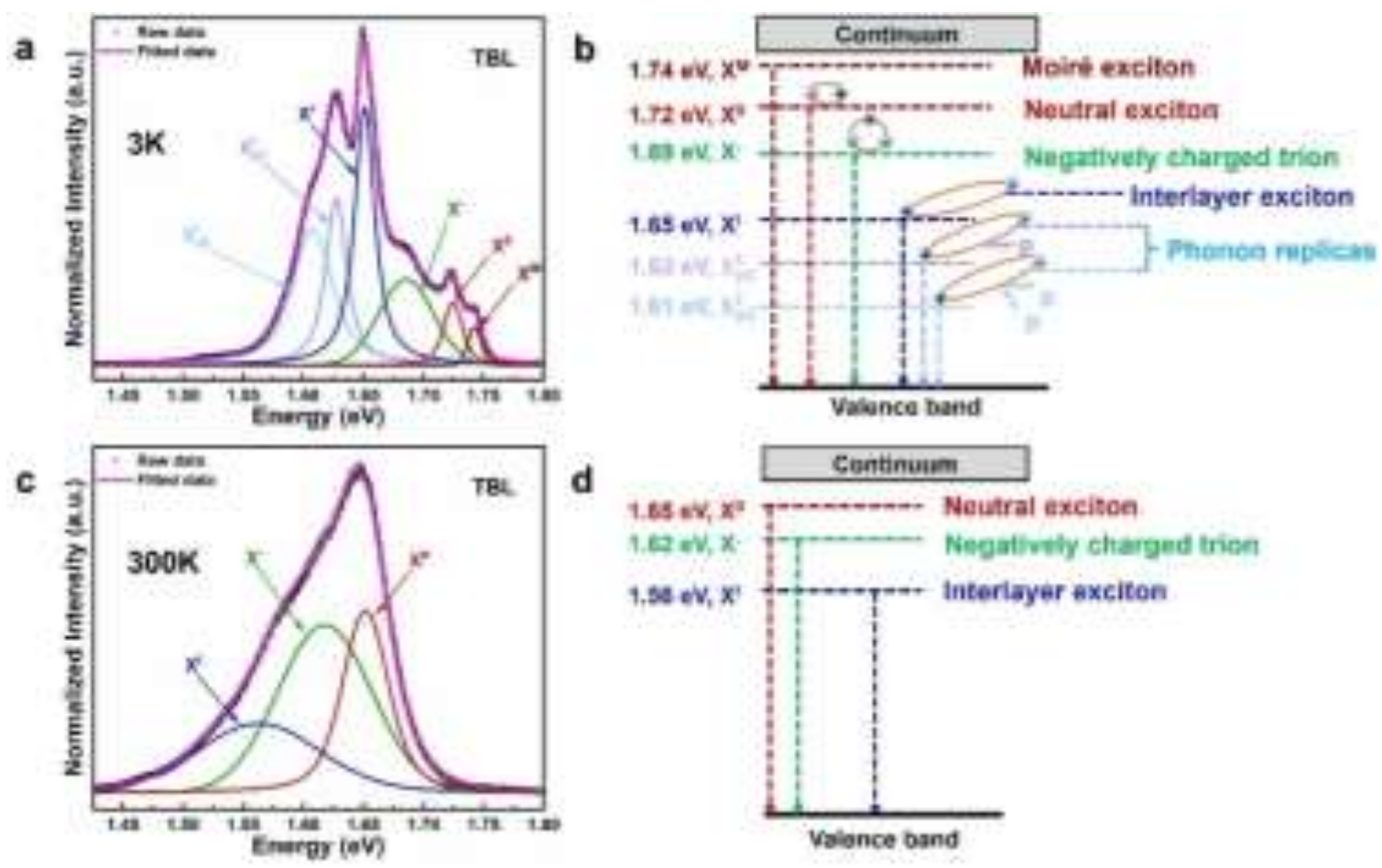

**Figure 5.** PL spectra and associated energy band diagrams at 3 K and 300 K for TBL. (a) PL spectra at 3 K, (b) Schematic of energy band diagram at 3 K. (c) PL spectrum at 300 K, (d) Schematic of energy band diagram at 300 K.

Figure 5a, b shows the proposed energy band diagram and schematic representation of the optical transitions of TBL, at 3 K. Considering the existence of $X^0$, $X^-$ and $X^I$ up to room temperature, we fitted the 300 K PL spectra with three peaks as shown in Figure 5c, observed at ~

1.65 eV, ~ 1.62 eV and ~ 1.56 eV, respectively. The Energy band diagram representing optical transitions at room temperature is shown in Figure 5d.[16] A clear observation of phonon-assisted excitonic emissions and moiré-potential induced emission with strong exciton-phonon coupling of interlayer excitons establishes precise control of twist angle as an innovative route for engineering excitonic emission in TMDs.

## Conclusion

In summary, we demonstrated the phononic and excitonic responses of $WSe_2$ across different stacking configurations (ML, NBL, ABL, and TBL) through Raman and PL spectroscopy. The twisting of the monolayers provides a modified periodic potential landscape that favors redistribution of electronic energy across multiple valleys and tailors phonon-assisted exciton dynamics in TBL encapsulated in hBN. This is well captured through the enhanced recombination efficiency and superimposition of moiré-potential induced emissions over the localized defect-bound emissions. Furthermore, the temperature dependent PL measurements confirm strong exciton-phonon interactions in TBL. Our findings position TBL as a tunable platform for quantum photonics and robust optoelectronic applications based on moiré-engineered exciton-phonon interactions in TMDs.

## Experimental Section

### Heterostructure Fabrication

The isolated layers of $WSe_2$ were mechanically exfoliated on a polydimethylsiloxane (PDMS) film from a well characterized bulk crystal. The monolayer was then transferred from PDMS onto a $SiO_2$/Si substrate, using the dry-transfer technique. To fabricate the twisted $WSe_2$ bilayer, the isolated sheet of $WSe_2$ was split into two parts using an optical fiber with an approximately 0.7 µm tip diameter.[24] Similarly, hBN flakes with a thickness of 10-15 nm were exfoliated on a $SiO_2$/Si substrate. The samples with twisted bilayers were then fabricated using the tear and stack technique. First, the top hBN layer ($hBN_T$) was picked up using the polypropylene carbonate (PPC) film placed on a PDMS stamp. A portion of $WSe_2$ monolayer was then picked up using the van der Waals forces between the $WSe_2$ monolayer and $hBN_T$ layer. The remaining portion of $WSe_2$ monolayer was then rotated to the desired angle and stacked with the pre-assembled $hBN_T/WSe_2$ layer. Another hBN layer ($hBN_B$) was then picked up and added at the bottom of the $hBN_T$/t-BL assembly. The hBN encapsulated sample was finally transferred onto a clean $SiO_2$/Si substrate. The fabricated samples were sequentially cleaned with chloroform and isopropyl alcohol (IPA) to eliminate surface contamination.

### Raman and PL Measurements

The Raman and PL spectroscopy measurements were carried out using a Horiba LabRAM HR Evolution spectrometer equipped with a ~ 532 nm excitation laser and a 1800 grooves/mm grating. Ultra-low frequency filters were used for low frequency Raman modes. For the PL measurements, a diffraction grating with 600 grooves/mm was used. The temperature-dependent measurements were performed using a Montana cryostat.

**Acknowledgements:** AS and AD acknowledges DST [DST/NPNST/AM/2025/31] for the funding support. MT, AT and AS thank IIT Mandi for the research facilities. AS acknowledges funding from Anusandhan National Research Foundation, New Delhi, India, under grant ANRF/ARG/2025/001724/PS.

## References

1. M. Yankowitz, S. Chen, H. Polshyn, et al., "Tuning Superconductivity in Twisted Bilayer Graphene," Science 363, no. 6431 (2019): 1059-1064, https://doi.org/10.1126/science.aav1910.
2. Z. Ma, Z. Hu, X. Zhao, et al., "Tunable Band Structures of Heterostructured Bilayers with Transition-Metal Dichalcogenide and MXene Monolayer," The Journal of Physical Chemistry C 118, no. 10 (2014): 5593-5599, https://doi.org/10.1021/jp500861n.
3. R. Wu, H. Zhang, H. Ma, et al., "Synthesis, Modulation, and Application of Two-Dimensional TMD Heterostructures," Chemical Reviews 124, no. 17 (2024): 10112-10191, https://doi.org/10.1021/acs.chemrev.4c00174.
4. M. M. Altaiary, E. Liu, C.-T. Liang, et al., "Electrically Switchable Intervalley Excitons with Strong Two-Phonon Scattering in Bilayer $WSe_2$," Nano Letters 22, no. 5 (2022): 1829-1835, https://doi.org/10.1021/acs.nanolett.1c01590.
5. M. Okada, A. Kutana, Y. Kureishi, et al., "Direct and Indirect Interlayer Excitons in a van der Waals Heterostructure of $hBN/WS_2/MoS_2/hBN$," ACS Nano 12, no. 3 (2018): 2498-2505, https://doi.org/10.1021/acsnano.7b08253.
6. C. Li, Q. Cao, F. Wang, et al., "Engineering Graphene and TMDs Based van der Waals Heterostructures for Photovoltaic and Photoelectrochemical Solar Energy Conversion," Chemical Society Reviews 47, no. 13 (2018): 4981-5037, https://doi.org/10.1039/C8CS00067K.
7. K. Tran, G. Moody, F. Wu, et al., "Evidence for Moiré Excitons in van der Waals Heterostructures," Nature 567, no. 7746 (2019): 71-75, https://doi.org/10.1038/s41586-019-0975-z.
8. A. Arora, M. Koperski, K. Nogajewski, et al., "Excitonic Resonances in Thin Films of $WSe_2$: From Monolayer to Bulk Material," Nanoscale 7, no. 23 (2015): 10421-10429, https://doi.org/10.1039/C5NR01536G.
9. S. Rai and A. Srivastava, "Low-Temperature Photoluminescence and Raman Study of Monolayer $WSe_2$ for Photocarrier Dynamics and Thermal Conductivity," Journal of Applied Physics 136, no. 15 (2024): 154301, https://doi.org/10.1063/5.0223732.
10. A. M. Jones, H. Yu, J. S. Ross, et al., "Spin–Layer Locking Effects in Optical Orientation of Exciton Spin in Bilayer $WSe_2$," Nature Physics 10, no. 2 (2014): 130-134, https://doi.org/10.1038/nphys2848.
11. A. Thomas, A. Singh, S. K. Sharma, and A. Soni, "Thermal Evolution of Excitons and Defect-Localized States in Monolayer 2H-$MoTe_2$," physica status solidi (RRL) – Rapid Research Letters 20, no. 1 (2026): e202500477, https://doi.org/10.1002/pssr.202500477.
12. P. Merkl, F. Mooshammer, S. Brem, et al., "Twist-Tailoring Coulomb Correlations in van der Waals Homobilayers," Nature Communications 11, no. 1 (2020): 2167, https://doi.org/10.1038/s41467-020-16069-z.
13. N. Gupta, S. Sachin, P. Kumari, S. Rani, and S. J. Ray, "Twistronics in Two-Dimensional Transition Metal Dichalcogenide (TMD)-Based van der Waals Interface," RSC Advances 14, no. 5 (2024): 2878-2888, https://doi.org/10.1039/D3RA06559F.
14. M. Angeli, G. R. Schleder, and E. Kaxiras, "Twistronics of Janus Transition Metal Dichalcogenide Bilayers," Physical Review B 106, no. 23 (2022): 235159, https://doi.org/10.1103/PhysRevB.106.235159.
15. W. Zhao, Z. Ghorannevis, L. Chu, et al., "Evolution of Electronic Structure in Atomically Thin Sheets of $WS_2$ and $WSe_2$," ACS Nano 7, no. 1 (2013): 791-797, https://doi.org/10.1021/nn305275h.
16. J. Pandey and A. Soni, "Unraveling Biexciton and Excitonic Excited States from Defect Bound States in Monolayer $MoS_2$," Applied Surface Science 463, no. (2019): 52-57, https://doi.org/10.1016/j.apsusc.2018.08.205.
17. C. Zhang, C. P. Chuu, X. Ren, et al., "Interlayer Couplings, Moiré Patterns, and 2D Electronic Superlattices in $MoS_2/WSe_2$ Hetero-Bilayers," Science Advances 3, no. 1 (2017): e1601459, https://doi.org/10.1126/sciadv.1601459.
18. J. Quan, L. Linhart, M.-L. Lin, et al., "Phonon Renormalization in Reconstructed $MoS_2$ Moiré Superlattices," Nature Materials 20, no. 8 (2021): 1100-1105, https://doi.org/10.1038/s41563-021-00960-1.

19. N. K. Singh and A. Soni, "Crystalline Anharmonicity and Ultralow Thermal Conductivity in Layered $Bi_2GeTe_4$ for Thermoelectric Applications," Applied Physics Letters 117, no. 12 (2020): 123901, https://doi.org/10.1063/5.0024651.
20. N. K. Singh, A. Kashyap, and A. Soni, "Ultralow Thermal Conductivity and Thermoelectric Properties of $Bi_4GeTe_7$ with an Intrinsic van der Waals Heterostructure," Applied Physics Letters 119, no. 22 (2021): 223903, https://doi.org/10.1063/5.0076785.
21. W. Zhao, R. M. Ribeiro, M. Toh, et al., "Origin of Indirect Optical Transitions in Few-Layer $MoS_2$, $WS_2$, and $WSe_2$," Nano Letters 13, no. 11 (2013): 5627-5634, https://doi.org/10.1021/nl403270k.
22. M. Van Winkle, I. M. Craig, S. Carr, et al., "Rotational and Dilational Reconstruction in Transition Metal Dichalcogenide Moiré Bilayers," Nat Commun 14, no. 1 (2023): 2989, https://doi.org/10.1038/s41467-023-38504-7.
23. D. Zhai, H. Yu, and W. Yao, "Twistronics and moiré superlattice physics in 2D transition metal dichalcogenides," Reports on Progress in Physics 88, no. 8 (2025): 084501, https://doi.org/10.1088/1361-6633/adefef.
24. K. P. Bera, D. Solanki, S. Mandal, et al., "Twist Angle-Dependent Phonon Hybridization in $WSe_2/WSe_2$ Homobilayer," ACS Nano 18, no. 35 (2024): 24379-24390, https://doi.org/10.1021/acsnano.4c06767.
25. T. Yan, X. Qiao, X. Liu, P. Tan, and X. Zhang, "Photoluminescence Properties and Exciton Dynamics in Monolayer $WSe_2$," Applied Physics Letters 105, no. 10 (2014): 101901, https://doi.org/10.1063/1.4895471.
26. K. Parto, S. I. Azzam, K. Banerjee, and G. Moody, "Defect and Strain Engineering of Monolayer $WSe_2$ Enables Site-Controlled Single-Photon Emission up to 150 K," Nature Communications 12, no. 1 (2021): 3585, https://doi.org/10.1038/s41467-021-23709-5.
27. G. Wang, X. Marie, L. Bouet, et al., "Exciton Dynamics in $WSe_2$ Bilayers," Applied Physics Letters 105, no. 18 (2014): 182105, https://doi.org/10.1063/1.4900945.
28. T. I. Andersen, G. Scuri, A. Sushko, et al., "Excitons in a Reconstructed Moiré Potential in Twisted $WSe_2/WSe_2$ Homobilayers," Nature Materials 20, no. 4 (2021): 480-487, https://doi.org/10.1038/s41563-020-00873-5.
29. M. H. Naik and M. Jain, "Ultraflatbands and Shear Solitons in Moiré Patterns of Twisted Bilayer Transition Metal Dichalcogenides," Physical Review Letters 121, no. 26 (2018): 266401, https://doi.org/10.1103/PhysRevLett.121.266401.
30. I. Maity, M. H. Naik, P. K. Maiti, H. R. Krishnamurthy, and M. Jain, "Phonons in Twisted Transition-Metal Dichalcogenide Bilayers: Ultrasoft Phasons and a Transition from a Superlubric to a Pinned Phase," Physical Review Research 2, no. 1 (2020): 013335, https://doi.org/10.1103/PhysRevResearch.2.013335.
31. M. R. Rosenberger, H.-J. Chuang, M. Phillips, et al., "Twist Angle-Dependent Atomic Reconstruction and Moiré Patterns in Transition Metal Dichalcogenide Heterostructures," ACS Nano 14, no. 4 (2020): 4550-4558, https://doi.org/10.1021/acsnano.0c00088.
32. J. Zhang, Z. Peng, A. Soni, et al., "Raman Spectroscopy of Few-Quintuple Layer Topological Insulator $Bi_2Se_3$ Nanoplatelets," Nano Letters 11, no. 6 (2011): 2407-2414, https://doi.org/10.1021/nl200773n.
33. W. Zhao, Z. Ghorannevis, K. K. Amara, et al., "Lattice Dynamics in Mono-and Few-Layer Sheets of $WS_2$ and $WSe_2$," Nanoscale 5, no. 20 (2013): 9677-9683, https://doi.org/10.1039/C3NR03052K.
34. S. Kim, K. Kim, J.-U. Lee, and H. Cheong, "Excitonic Resonance Effects and Davydov Splitting in Circularly Polarized Raman Spectra of Few-Layer $WSe_2$," 2D Materials 4, no. 4 (2017): 045002, https://doi.org/10.1088/2053-1583/aa8312.
35. P. Tonndorf, R. Schmidt, R. Schneider, et al., "Single-Photon Emission from Localized Excitons in an Atomically Thin Semiconductor," Optica 2, no. 4 (2015): 347-352, https://doi.org/10.1364/OPTICA.2.000347.
36. B. Wu, H. Zheng, S. Li, et al., "Evidence for Moiré Intralayer Excitons in Twisted $WSe_2/WSe_2$ Homobilayer Superlattices," Light: Science & Applications 11, no. 1 (2022): 166, https://doi.org/10.1038/s41377-022-00854-0.
37. A. Laturia, M. L. Van de Put, and W. G. Vandenberghe, "Dielectric Properties of Hexagonal Boron Nitride and Transition Metal Dichalcogenides: from Monolayer to Bulk," npj 2D Materials and Applications 2, no. 1 (2018): 6, https://doi.org/10.1038/s41699-018-0050-x.
38. C. Lee, X. Wei, J. W. Kysar, and J. Hone, "Measurement of the Elastic Properties and Intrinsic Strength of Monolayer Graphene," Science 321, no. 5887 (2008): 385-388, https://doi.org/10.1126/science.1157996.
39. L. Fang, H. Chen, X. Yuan, et al., "Quick Optical Identification of the Defect Formation in Monolayer $WSe_2$ for Growth Optimization," Nanoscale Research Letters 14, no. 1 (2019): 274, https://doi.org/10.1186/s11671-019-3110-z.
40. L. Wang, E.-M. Shih, A. Ghiotto, et al., "Correlated Electronic Phases in Twisted Bilayer Transition Metal Dichalcogenides," Nature Materials 19, no. 8 (2020): 861-866, https://doi.org/10.1038/s41563-020-0708-6.

41. Z. Li, J. Förste, K. Watanabe, et al., "Stacking-Dependent Exciton Multiplicity in $WSe_2$ Bilayers," Physical Review B 106, no. 4 (2022): 045411, https://doi.org/10.1103/PhysRevB.106.045411.
42. F. Wu, T. Lovorn, E. Tutuc, I. Martin, and A. H. MacDonald, "Topological Insulators in Twisted Transition Metal Dichalcogenide Homobilayers," Physical Review Letters 122, no. 8 (2019): 086402, https://doi.org/10.1103/PhysRevLett.122.086402.
43. R. Debnath, S. Sett, S. Kundu, et al., "Tuning Exciton Complexes in Twisted Bilayer $WSe_2$ at Intermediate Misorientation," Physical Review B 106, no. 12 (2022): 125409, https://doi.org/10.1103/PhysRevB.106.125409.
44. P. K. Barman, P. Upadhyay, R. Rajarapu, et al., "Twist-Dependent Tuning of Excitonic Emissions in Bilayer $WSe_2$," ACS Omega 7, no. 7 (2022): 6412-6418, https://doi.org/10.1021/acsomega.1c07219.
45. A. Kumar, D. Solanki, K. Watanabe, et al., "Interlayer Phonon Coupling and Enhanced Electron–Phonon Interactions in Doubly Aligned hBN/Graphene/hBN Heterostructures," ACS Nano 19, no. 17 (2025): 16415-16423, https://doi.org/10.1021/acsnano.4c17152.
46. S. Brem, C. Linderälv, P. Erhart, and E. Malic, "Tunable Phases of Moiré Excitons in van der Waals Heterostructures," Nano Letters 20, no. 12 (2020): 8534-8540, https://doi.org/10.1021/acs.nanolett.0c03019.
47. R. Narciso Pedrosa, C. E. P. Villegas, A. Reily Rocha, R. G. Amorim, and W. Scopel, "Interlayer Excitons and Radiative Lifetimes in $MoSe_2$/SeWS Bilayers: Implications for Light-Emitting Diodes," ACS Applied Nano Materials 8, no. 10 (2025): 5051-5058, https://doi.org/10.1021/acsanm.4c06994.
48. B. Miller, A. Steinhoff, B. Pano, et al., "Long-Lived Direct and Indirect Interlayer Excitons in van der Waals Heterostructures," Nano Letters 17, no. 9 (2017): 5229-5237, https://doi.org/10.1021/acs.nanolett.7b01304.
49. E. Liu, J. van Baren, T. Taniguchi, et al., "Magnetophotoluminescence of Exciton Rydberg States in Monolayer $WSe_2$," Physical Review B 99, no. 20 (2019): 205420, https://doi.org/10.1103/PhysRevB.99.205420.
50. A. Delhomme, G. Butseraen, B. Zheng, et al., "Magneto-Spectroscopy of Exciton Rydberg States in a CVD Grown $WSe_2$ Monolayer," Applied Physics Letters 114, no. 23 (2019): 232104, https://doi.org/10.1063/1.5095573.
51. T. Wang, Z. Li, Y. Li, et al., "Giant Valley-Polarized Rydberg Excitons in Monolayer $WSe_2$ Revealed by Magneto-Photocurrent Spectroscopy," Nano Letters 20, no. 10 (2020): 7635-7641, https://doi.org/10.1021/acs.nanolett.0c03167.
52. N. Saigal and S. Ghosh, "Evidence for Two Distinct Defect Related Luminescence Features in Monolayer $MoS_2$," Applied Physics Letters 109, no. 12 (2016): 122105, https://doi.org/10.1063/1.4963133.
53. S. Tongay, J. Suh, C. Ataca, et al., "Defects Activated Photoluminescence in Two-Dimensional Semiconductors: Interplay Between Bound, Charged, and Free Excitons," Scientific Reports 3, no. (2013): https://doi.org/10.1038/srep02657.
54. B. Xie and J. Liu, "Lattice Distortions, Moiré Phonons, and Relaxed Electronic Band Structures in Magic-Angle Twisted Bilayer Graphene," Physical Review B 108, no. 9 (2023): 094115, https://doi.org/10.1103/PhysRevB.108.094115.
55. C. Chen, K. P. Nuckolls, S. Ding, et al., "Strong Electron–Phonon Coupling in Magic-Angle Twisted Bilayer Graphene," Nature 636, no. 8042 (2024): 342-347, https://doi.org/10.1038/s41586-024-08227-w.
56. Z. Li, T. Wang, C. Jin, et al., "Emerging Photoluminescence from the Dark-Exciton Phonon Replica in Monolayer $WSe_2$," Nature Communications 10, no. 1 (2019): 2469, https://doi.org/10.1038/s41467-019-10477-6.
57. C. Blaga, Á. Labordet Álvarez, A. Balgarkashi, et al., "Unveiling the Complex Phonon Nature and Phonon Cascades in 1L to 5L $WSe_2$ Using Multiwavelength Excitation Raman Scattering," Nanoscale Advances 6, no. 18 (2024): 4591-4603, https://doi.org/10.1039/d4na00399c.
58. H. Sahin, S. Tongay, S. Horzum, et al., "Anomalous Raman spectra and thickness-dependent electronic properties of $WSe_2$," Physical Review B 87, no. 16 (2013): 165409, https://doi.org/10.1103/PhysRevB.87.165409.
59. J. Kopaczek, S. Zelewski, K. Yumigeta, et al., "Temperature Dependence of the Indirect Gap and the Direct Optical Transitions at the High-Symmetry Point of the Brillouin Zone and Band Nesting in $MoS_2$, $MoSe_2$, $MoTe_2$, $WS_2$, and $WSe_2$ Crystals," The Journal of Physical Chemistry C 126, no. 12 (2022): 5665-5674, https://doi.org/10.1021/acs.jpcc.2c01044.

## Supporting Information

### Moiré-Engineered Excitonic Landscape and Phonon-Mediated Recombination in Twisted $WSe_2$ Bilayers

*Memansa Thapa[1], Aksa Thomas[1], Jayalekshmi U. J.[2,3], Krishna Prasad Bera[4], Darshit Solanki[4], Kenji Watanabe[5], Takashi Taniguchi[5], Ajay Kumar Shukla[2,3], Anindya Das[4], and Ajay Soni[1]**

[1]School of Physical Sciences, Indian Institute of Technology Mandi, Mandi 175005, Himachal Pradesh, India
[2]CSIR-National Physical Laboratory, Dr. K. S. Krishnan Marg, New Delhi, 110012, India
[3]Academy of Scientific and Innovative Research (AcSIR), Ghaziabad, 201002, India
[4]Department of Physics, Indian Institute of Science, Bangalore 560012, Karnataka, India
[5]National Institute of Material Science, 1-1 Namiki, Tsukuba 305-0044, Japan
[*]*Email for Correspondence:* *ajay@iitmandi.ac.in*

The Supporting Information provides detailed characterizations and photoluminescence (PL) analyses that complement the main manuscript. Specifically, it includes, (i) the temperature-dependent Raman spectroscopy for TBL, highlighting the thermal evolution of phonons in Figure S1; (ii) comparison of the room-temperature PL spectra of all $WSe_2$ samples used in the manuscript in Figure S2; (iii) relative integrated PL intensity plot for TBL in Figure S3; and (iv) fitting comparison of the PL analyzed using both the Varshni and O'Donnell models for ML (Figure S4), NBL (Figure S5) and TBL (Figure S6). These complementary datasets and model fittings provide a comprehensive and self-consistent understanding of the temperature-dependent PL behavior discussed in the main manuscript.

### S1. Temperature-dependent Raman spectroscopy of TBL $WSe_2$.

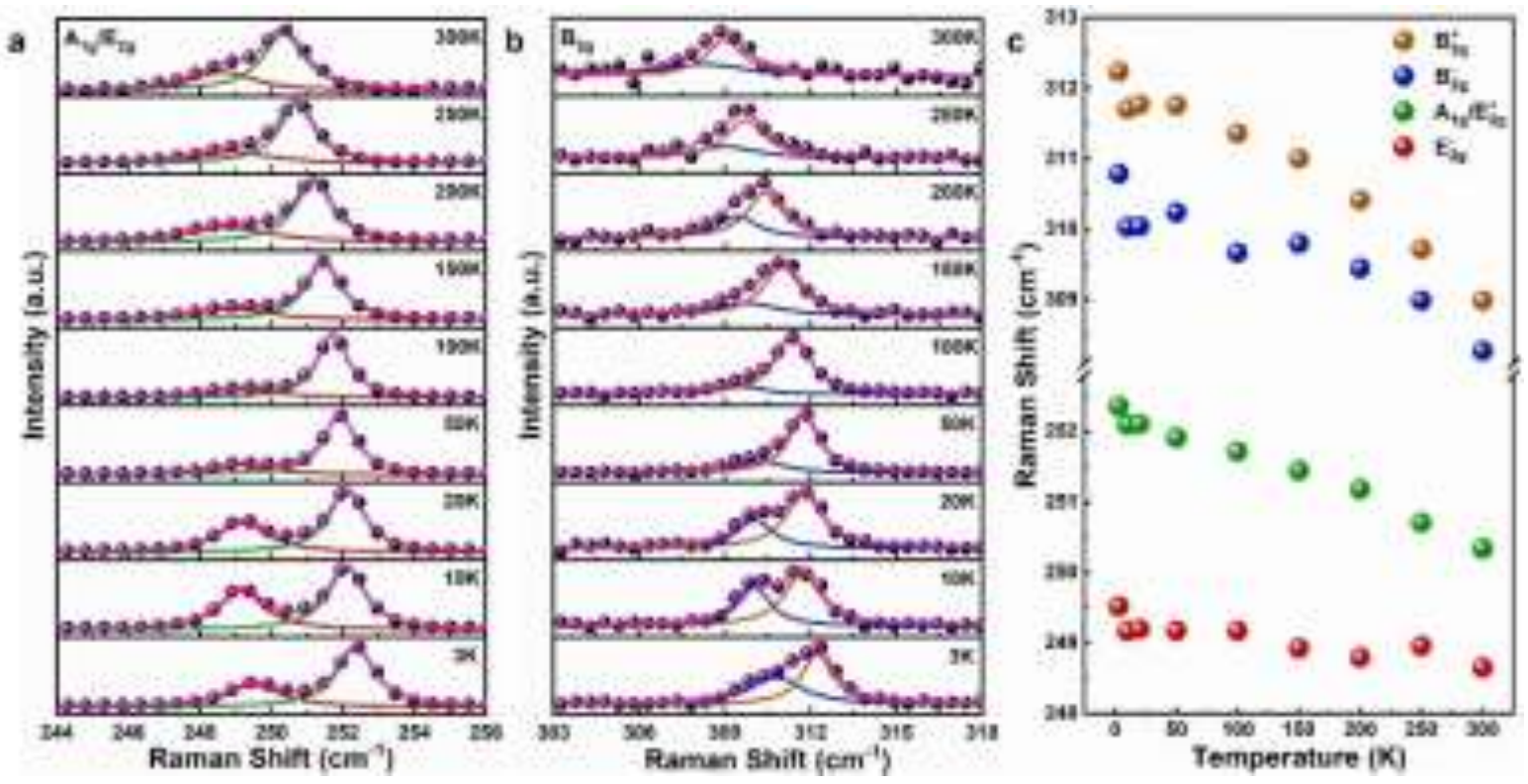

**Figure S1.** Temperature-dependent Raman spectra and splitting of phonons for TBL. Raman spectra of (a) $A_{1g}$ / $E_{2g}$ mode; and (b) $B_{2g}$ mode; (c) Peak positions vs temperature for $E_{2g}^{-}$, $A_{1g}$ / $E_{2g}^{+}$, $B_{2g}^{-}$, and $B_{2g}^{+}$ modes.

Figure S1 shows the temperature-dependent evolution of the Raman modes of TBL. The Raman spectra in the frequency range 244 - 256 cm$^{-1}$ (Figure S1a), displays the splitting of $A_{1g}$ / $E_{2g}$ mode into a doublet, $E_{2g}^{-}$ (red solid lines) and $A_{1g}$ / $E_{2g}^{+}$ (green solid lines). Figure S1b shows the thermal evolution of the interlayer coupling mode $B_{2g}$ arising from the out of plane displacement of two Se atoms with respect to the W atom. The $B_{2g}$ mode also splits into a doublet, $B_{2g}^{-}$ (blue solid lines) and $B_{2g}^{+}$ (orange solid lines). The observed splitting of the Raman modes is consistent with the previously reported literature by Bera *et al.,*[1] where the splitting is attributed to the phonon hybridization, arising from the moiré potential formed due to the twisting of the two layers.[1,2] Figure S1c shows the softening of the phonon as a redshift in the peak positions with respect to the temperature. The consistency of our data, with the previously reported literature, not only confirms the presence of phonon hybridization in TBL, but also validates the 2º twisting angle through strong Raman signatures.

**S2. Room Temperature PL spectra and mapping of various heterostructures.**

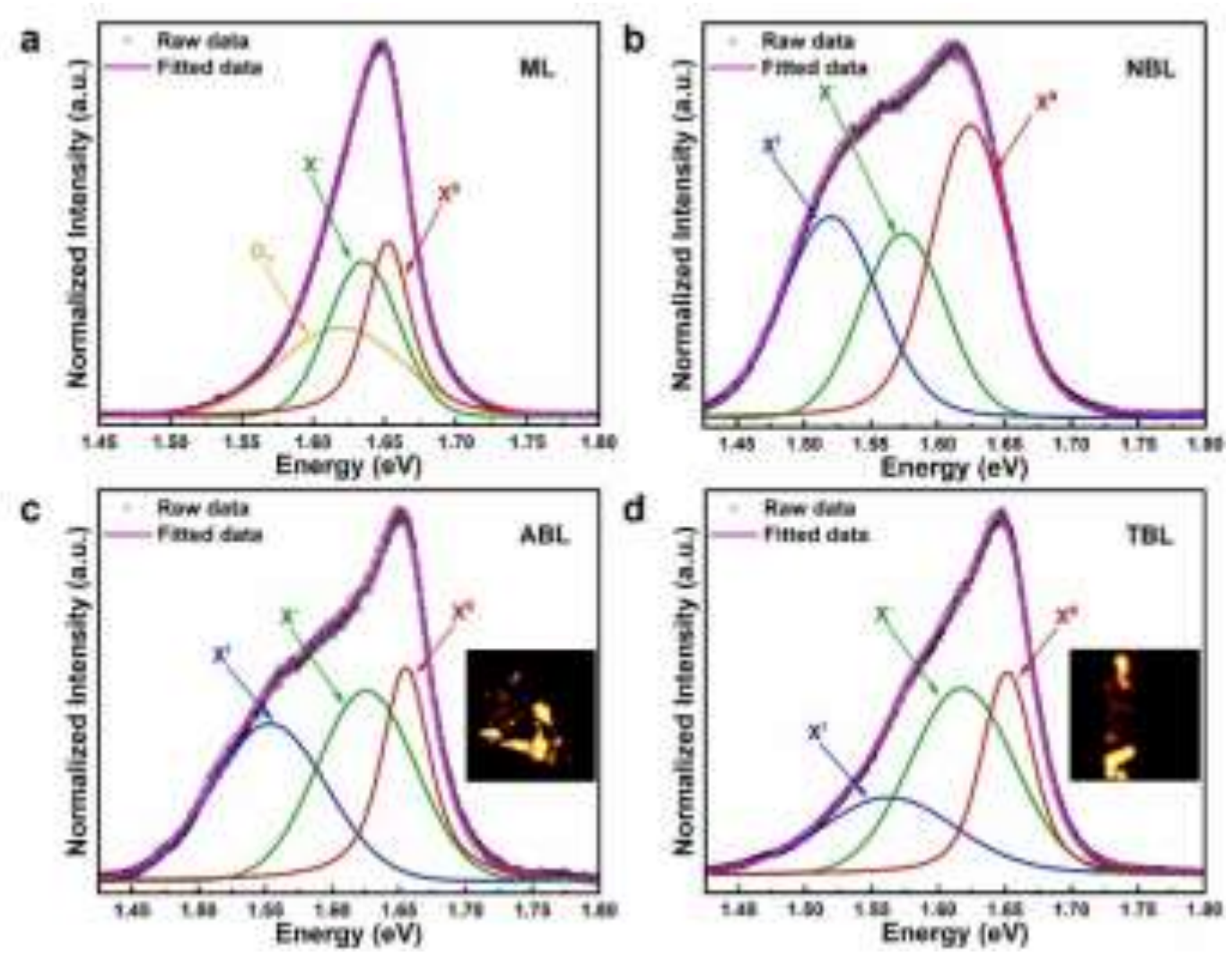

**Figure S2.** Normalized PL spectra at 300 K for various stacking layers of $WSe_2$. (a) monolayer (ML), (b) natural bilayer (NBL), (c) artificial bilayer fabricated with twist angle, $\theta \sim 60^{\circ}$ (ABL), and (d) twisted bilayer with $\theta \sim 2^{\circ}$ (TBL). The insets in panels (c) and (d) show the corresponding PL intensity mappings of the twisted and monolayer regions, confirming the overlaps and formation of heterostructures.

Figure S2 presents the room-temperature PL of various $WSe_2$ layers in different stacking configurations. The monolayer $WSe_2$ exhibits a strong emission, which is a characteristic of direct bandgap (Figure S2a), the natural bilayer has a reduced PL intensity indicating the transition to an indirect bandgap (Figure S2b). Similar PL features are observed for the artificially stacked bilayer encapsulated with hBN, (Figure S2c). The artificial bilayer with ~ 60º twist angle is expected to behave as the natural bilayer with 2H stacking. Figure S2d displays the PL response of a 2º twisted bilayer. The insets in Figure S2c, d show the corresponding PL intensity mappings of the measured regions, with the bright areas revealing the emissions from the monolayer domains, while the suppressed areas reveal the bilayer regions.[3,4]

## S3. Excitonic complexes in TBL.

**Table S1.** Excitonic complexes in TBL $WSe_2$.

| Excitonic complex | This paper | Literature |
|---|---|---|
| $X^0$ | ~ 1.72 eV | ~ 1.735 eV (20º t-$WSe_2$)[5],<br>~ 1.733 eV (monolayer)[6],<br>~ 1.703 eV (monolayer)[7] |
| $X^-$ | ~ 1.69 eV | ~ 1.703 eV (20º t-$WSe_2$)[5],<br>~ 1.697 eV (monolayer)[6],<br>~ 1.673 eV (monolayer) |
| $X^I$ | ~ 1.65 eV | ~ 1.675 eV (20º t-$WSe_2$)[5],<br>~ 1.59 eV (21º t-$WSe_2$)[8], |
| $X^I_p$ | ~ 1.63 eV, ~ 1.61 eV | ~ 1.660 eV, ~ 1.647 eV (20º t-$WSe_2$)[5],<br>1.5eV-1.7 eV[9] |

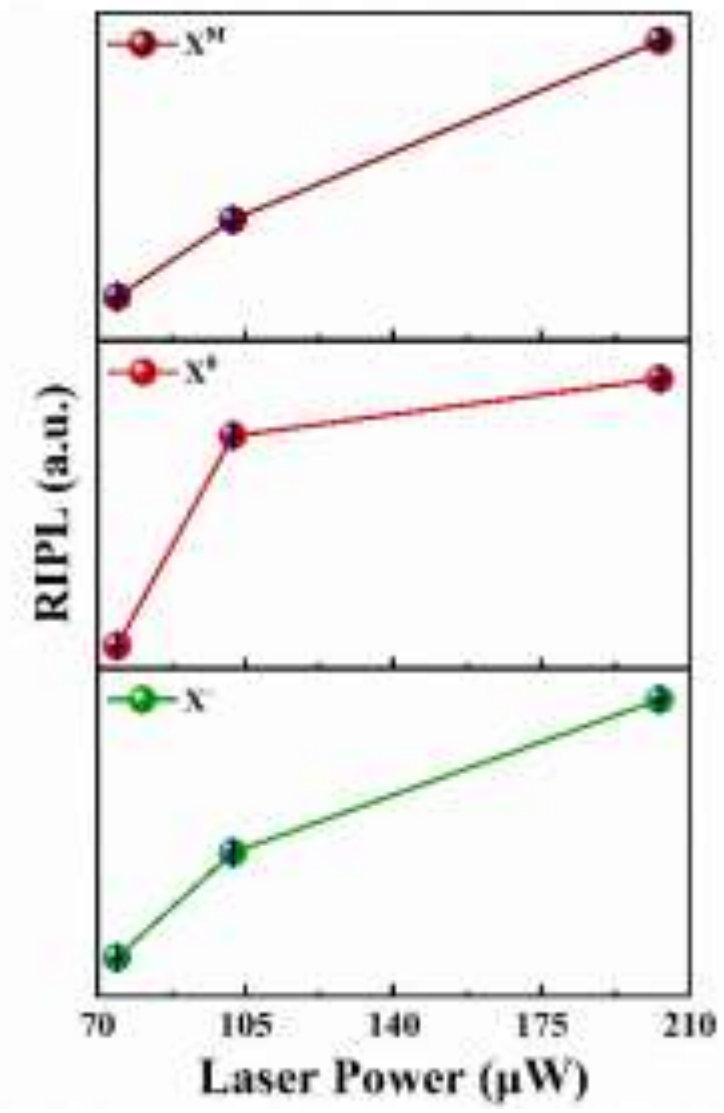

**Figure S3.** Relative integrated PL intensity with laser excitation for moiré exciton ($X^M$), neutral exciton ($X^0$), and trion ($X^-$) in the twisted bilayer $WSe_2$.

**S4. Varshni and O'Donnell fitting analysis for various stacked layer.**

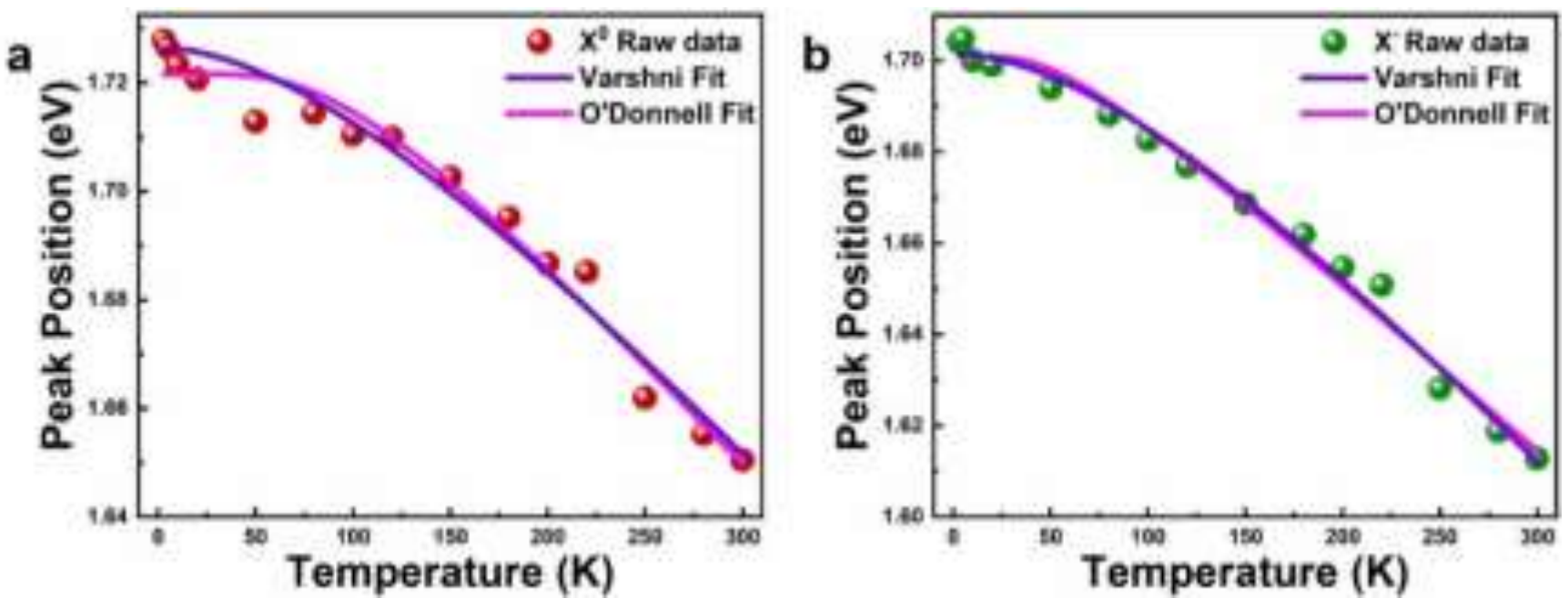

**Figure S4.** Varshni equation and O′Donnell model fitting of temperature-dependent shift of (a) neutral exciton ($X^0$) and (b) trion ($X^-$) in monolayer $WSe_2$.

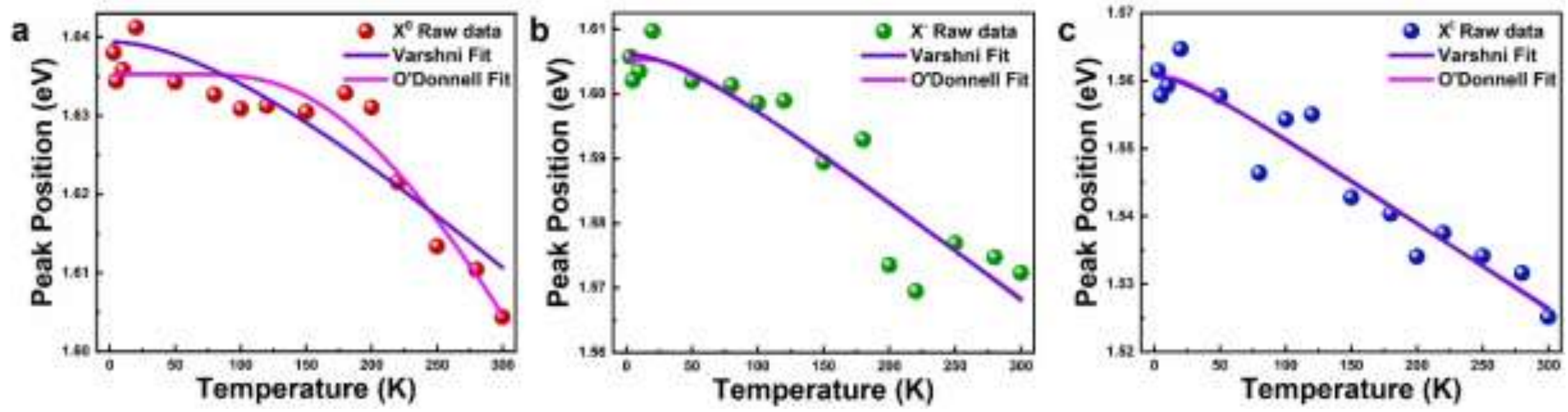

**Figure S5.** Varshni equation and O′Donnell model fitting of temperature-dependent shift of (a) neutral exciton ($X^0$), (b) trion ($X^-$), and (c) indirect exciton ($X^I$) in natural bilayer $WSe_2$.

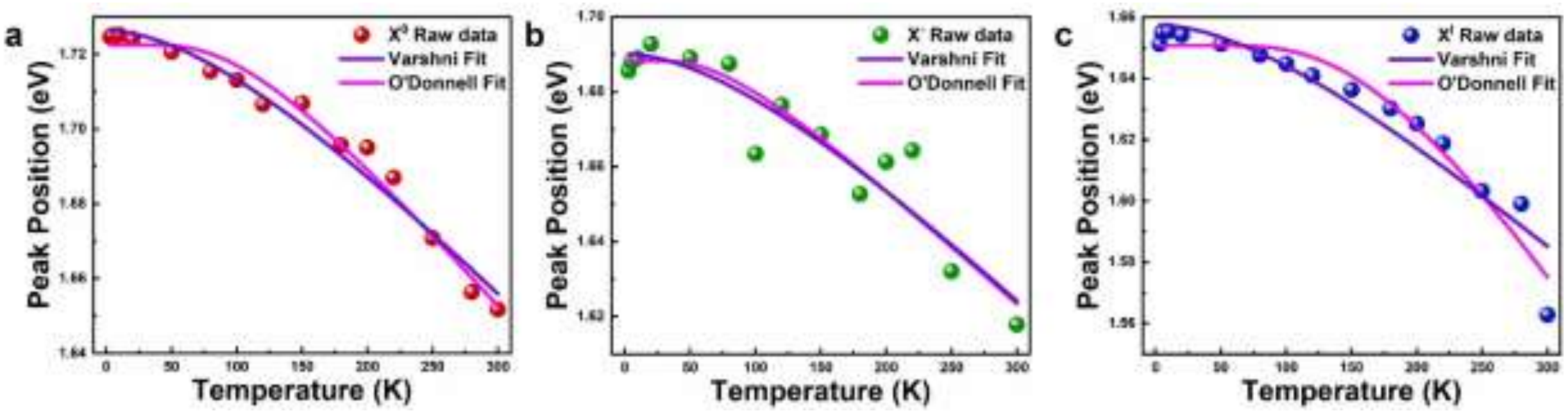

**Figure S6.** Varshni equation and O′Donnell model fitting of temperature-dependent shift of (a) neutral exciton ($X^0$), (b) trion ($X^-$), and (c) indirect exciton ($X^I$) in twisted bilayer $WSe_2$.

**Table S2.** Varshni equation fitting parameters ($\alpha$ and $\beta$) for Temperature-Dependent exciton Peaks in $WSe_2$ systems.

| Layer configuration | Excitonic complex | $\alpha$ (meV/K) | $\beta$ (K) |
|---|---|---|---|
| ML | $X^0$ | 0.405 | 200 |
| | $X^-$ | 0.494 | 200 |
| NBL | $X^0$ | 0.159 | 199 |
| | $X^-$ | 0.158 | 77 |
| | $X^I$ | 0.126 | 40 |
| TBL | $X^0$ | 0.391 | 200 |
| | $X^-$ | 0.366 | 200 |
| | $X^I$ | 0.402 | 200 |

The Varshni equation is a widely used empirical model to analyze the temperature dependence of the optical bandgap.[10] As the temperature increases, the lattice expands, increasing the vibrational energy, leading to a reduction in the bandgap energy. This temperature dependent behavior of the bandgap is a direct consequence of changes in the relative positions of the conduction and valence bands. These shifts are primarily governed by the exciton-phonon interactions, phonons and the thermal expansion of the lattice.[11,12] The temperature variation of the bandgap energy $E_g(T)$ is quantitatively given by the Varshni equation, $E_g(T) = E_g(0) - \frac{\alpha T^2}{T+\beta}$, where $E_g(0)$ denotes the intrinsic bandgap energy at absolute zero temperature and the parameter $\alpha$ describes the band gap change with temperature due to thermal expansion of the lattice, and the parameter $\beta$ is related to the Debye temperature. The extracted Varshni parameters further give insights into the origin of bandgap renormalization in TBL. The value of $\alpha$, which lies between those for monolayer and bilayer systems reflect the modified lattice environment in TBL, suggesting that the interlayer coupling and moiré-induced structural reconstruction leads to a redistribution of strain and bonding strengths, thereby effectively tuning the strength of exciton-phonon interactions, which are studied using the O'Donnell equation. The parameter $\beta$ ~ 200 K corresponds to a characteristic phonon temperature scale and is of the same order as the effective Debye temperature of the $WSe_2$ system, indicating that phonon populations relevant to bandgap renormalization become significant within the probed temperature range.

**Table S3.** O'Donnell equation fitting parameters ($S$ and $\hbar\omega$) for Temperature-Dependent exciton Peaks in $WSe_2$ systems.

| Layer configuration | Excitonic complex | $S$ | $\hbar\omega$ (meV) |
|---|---|---|---|
| ML | $X^0$ | 2.316 | 24.5 |
| | $X^-$ | 2.183 | 13.2 |
| NBL | $X^0$ | 2.295 | 23.6 |
| | $X^-$ | 0.868 | 9.7 |
| | $X^I$ | 0.719 | 4.6 |
| TBL | $X^0$ | 2.391 | 27.4 |
| | $X^-$ | 1.880 | 19.7 |
| | $X^I$ | 4.270 | 48.2 |

## References

1. K. P. Bera, D. Solanki, S. Mandal, et al., "Twist Angle-Dependent Phonon Hybridization in $WSe_2$/$WSe_2$ Homobilayer," ACS Nano 18, no. 35 (2024): 24379-24390, https://doi.org/10.1021/acsnano.4c06767.
2. J. Quan, L. Linhart, M.-L. Lin, et al., "Phonon Renormalization in Reconstructed $MoS_2$ Moiré Superlattices," Nature Materials 20, no. 8 (2021): 1100-1105, https://doi.org/10.1038/s41563-021-00960-1.
3. P. K. Barman, P. Upadhyay, R. Rajarapu, et al., "Twist-Dependent Tuning of Excitonic Emissions in Bilayer $WSe_2$," ACS Omega 7, no. 7 (2022): 6412-6418, https://doi.org/10.1021/acsomega.1c07219.
4. G. Wang, X. Marie, L. Bouet, et al., "Exciton Dynamics in $WSe_2$ Bilayers," Applied Physics Letters 105, no. 18 (2014): 182105, https://doi.org/10.1063/1.4900945.
5. R. Debnath, S. Sett, S. Kundu, et al., "Tuning Exciton Complexes in Twisted Bilayer $WSe_2$ at Intermediate Misorientation," Physical Review B 106, no. 12 (2022): 125409, https://doi.org/10.1103/PhysRevB.106.125409.
6. Z. Li, T. Wang, C. Jin, et al., "Momentum-Dark Intervalley Exciton in Monolayer Tungsten Diselenide Brightened via Chiral Phonon," ACS Nano 13, no. 12 (2019): 14107-14113, https://doi.org/10.1021/acsnano.9b06682.
7. S.-Y. Chen, T. Goldstein, T. Taniguchi, K. Watanabe, and J. Yan, "Coulomb-Bound Four- and Five-Particle Intervalley States in an Atomically-Thin Semiconductor," Nature Communications 9, no. 1 (2018): 3717, https://doi.org/10.1038/s41467-018-05558-x.
8. P. Merkl, F. Mooshammer, S. Brem, et al., "Twist-Tailoring Coulomb Correlations in van der Waals Homobilayers," Nature Communications 11, no. 1 (2020): 2167, https://doi.org/10.1038/s41467-020-16069-z.
9. M. M. Altaiary, E. Liu, C.-T. Liang, et al., "Electrically Switchable Intervalley Excitons with Strong Two-Phonon Scattering in Bilayer $WSe_2$," Nano Letters 22, no. 5 (2022): 1829-1835, https://doi.org/10.1021/acs.nanolett.1c01590.
10. Y. P. Varshni, "Temperature Dependence of the Energy Gap in Semiconductors," Physica 34, no. 1 (1967): 149-154, https://doi.org/10.1016/0031-8914(67)90062-6.
11. J. Kopaczek, S. Zelewski, K. Yumigeta, et al., "Temperature Dependence of the Indirect Gap and the Direct Optical Transitions at the High-Symmetry Point of the Brillouin Zone and Band Nesting in $MoS_2$, $MoSe_2$, $MoTe_2$, $WS_2$, and $WSe_2$ Crystals," The Journal of Physical Chemistry C 126, no. 12 (2022): 5665-5674, https://doi.org/10.1021/acs.jpcc.2c01044.
12. C.-C. Li, M. Gong, X.-D. Chen, et al., "Temperature Dependent Energy Gap Shifts of Single Color Center in Diamond Based on Modified Varshni Equation," Diamond and Related Materials 74, no. (2017): 119-124, https://doi.org/10.1016/j.diamond.2017.03.002.